# A 3D dynamical biomechanical tongue model to study speech motor control

Jean-Michel Gérard[1], Reiner Wilhelms-Tricarico[2], Pascal Perrier[1] & Yohan Payan[3]

[1]Institut de la Communication Parlée, UMR CNRS 5009, INPG & Université Stendhal, Grenoble, France
[2]Speech Communication Group, R.L.E., Massachusetts Institute of Technology, Cambridge, Massachusetts, USA
[3]Laboratoire TIMC, CNRS, Université Joseph Fourier, Grenoble, France

Contact Author:     Jean-Michel Gérard
                    Institut de la Communication Parlée
                    INPG
                    46 Avenue Félix Viallet
                    38031 Grenoble Cédex 01
                    France
                    Fax: 33 + (4)76 57 47 10
                    E-mail: gerard@icp.inpg.fr

# Abstract

A 3D biomechanical dynamical model of human tongue is presented, that is elaborated in the aim to test hypotheses about speech motor control. Tissue elastic properties are accounted for in Finite Element Modeling (FEM). The FEM mesh was designed in order to facilitate the implementation of muscle arrangement within the tongue. Therefore, its structure was determined on the basis of accurate anatomical data about the tongue. Mechanically, the hypothesis of hyperelasticity was adopted with the Mooney-Rivlin formulation of the strain energy function. Muscles are modeled as general force generators that act on anatomically specified sets of nodes of the FEM structure. The 8 muscles that are known to be largely involved in the production of basic speech movements are modeled. The model and the solving of the Lagrangian equations of movement are implemented using the ANSYS$^{TM}$ software. Simulations of the influence of muscle activations onto the tongue shape are presented and analyzed.

# I. Introduction

The study of human motor control implies to collect and to analyze a large number of kinematic data measured on the final effectors, such as the hand, the arm or the legs. These observations are considered to characterize the behavior of the peripheral motor system during different tasks and under variable conditions, and their interpretation contributed largely to the elaboration of major theories of motor control.

Nevertheless, a major methodological issue of this approach is associated with the fact, that peripheral signals are, actually, the result of a, often complicated, interaction between the motor control system and the peripheral apparatus. Both systems do contribute to shape the measured signals, in order to properly infer motor control strategies form these signal, it is important to assess the possible contribution of the physical systems. A very good illustration of this statement can be found in the controversy between Kawato and colleagues (Gomi & Kawato, 1996) [14] and Ostry and colleagues (Gribble *et al.*, 1998) [15] about Feldman's Equilibrium Point Hypothesis for motor control: using different biomechanical models of the arm they end up with opposite conclusions, while studying from the same kind of experimental data.

In this framework, an interesting approach, which is complementary to the experimental one, consists in building up physical models of the peripheral apparatus, to control them according to specific motor control model, and to compare the obtained simulations with experimental data. This is our approach to study speech motor control. This is why we developed in the last decade biomechanical models of the tongue and of the jaw (Wilhelms-Tricarico, 1995 [39]; Payan & Perrier, 1997 [28]; Sanguineti *et al.*, 1997 [33] ; Sanguineti *et al.*, 1998 [34]; Perrier *et al.*, to appear [32]).

Speech is a motor task, during which speaker communicate a message using temporal variation of an acoustic signal that is shaped by orofacial gestures, while listeners perceive this message using both the auditory perception of the acoustic signal and the vision of the articulatory gestures. The semiotic characteristics of these tasks make the study of its motor control very complex, since the task itself cannot be simply characterized in a physical space. The information can be retrieved via the combination of temporal and spectral properties, both in the auditory and in the visual domains. This is why biomechanical models used to study speech motor control and to study speech production have to be dynamical models: a realistic time variation of the simulated gestures is a necessity, if we want to test our models via their effects on the auditory and visual perception of the associated physical signals.

Contrary to the gestures mostly studied in motor control research, the tongue is moving in a closed space, the vocal tract. Consequently, during its movements the tongue is often in contact with external structures, such as the teeth, the palate or the pharyngeal walls (Stone, 1995 [35]; Fuchs *et al.*, 2001 [12]), which in turn contribute to the measured kinematic properties. In order to give a fair account of this factor, a three dimensional description of the tongue is required.

Based on these statements, we recently developed a 3D dynamical biomechanical model of the human tongue. This paper presents the structure of this model and some results about the modelled influence of the main tongue muscles onto the shape of this articulator.

# II. Design of the Finite Element Mesh

The design of the mesh was guided by the basic double requirement to build up a 3D geometrical structure that, on the one hand, could match as closely as possible the muscle morphology within the tongue, and, on the other hand, could be applicable to computational simulations based on finite element method (FEM) (Wilhelms-Tricarico 1995) [39].

A clear separation between muscles is usually not possible, since different muscles fibers often interleave with each other. Nevertheless, a schematic separation can be proposed, which is acceptable in the framework of a realistic biomechanical description. In the present study, the data set provided by the Visible Human Project® for a female subject was utilized to obtain quantitative measurements of tongue muscle anatomy and to incorporate them directly into an appropriate data structure from which a finite element model can be derived. The sets are series of digitized images of cryo-sections of the whole human body. To analyze these images, a software component was built that makes it possible to simultaneously display several arbitrarily oriented cross sections of the data block. This software provides visualization of the data together with tools to specify and edit geometrical objects such as points, lines and geometric solids in direct reference to any of a number of selected cross sectional planes. By these means, three-dimensional coordinates of points inside the structure can be directly obtained to specify and edit the shape of the model (Wilhelms-Tricarico, 2000) [40]. In order to associate the different elements of the FEM structure with anatomical tongue components, anatomy books, such as Pernkopf (1980) [31] and Netter (1989) [24], were used together with Miyawaki's data on tongue musculature (Miyawaki, 1974) [22].

For some muscles for which the actual distribution within the tongue is still a matter of debates and investigations, and for which the analysis of the Visible Human Project® data set did not yield an undisputable interpretation, Takemoto's (2001) [37] recent and very comprehensive functional presentation of tongue anatomy served as the main guide-line for obtaining muscle representations.

### II.1 *Some Basics about Tongue Anatomy*

A detailed anatomical study of the tongue musculature has been described in Takemoto (2001) [37]. Thus, the description given here will only address functional aspects (Perkins and Kent, 1986) [30]. Most of the muscles are paired, with one on each side of the midsagittal plane; however in the following description, their names are given in singular form.

Among the nine muscles that act on the tongue structure, there are five extrinsic muscles that originate on structures external to the tongue (mostly bony structures) and insert into the tongue: the *genioglossus*, the *styloglossus*, the *hyoglossus*, the *geniohyoid* and the *palatoglossus*. The influence of the *genioglossus* can be separated into three main actions: contraction of its posterior fibers produces a forward and upward movement of the tongue body, while its anterior fibers pull the anterior portion of the tongue downward; and the recruitment of its medium part flattens the tongue in the velar region. The *styloglossus* raises and retracts the tongue, causing a bunching of the dorsum in the velar region. The *hyoglossus* retracts and lowers the tongue body. The *geniohyoid* is the main component of the mouth floor. Through its action on the hyoid bone, it contributes to the displacement of the tongue root, and determines the protrusion of the tongue. The *palatoglossus* is a thin muscle attached on the velum on one end, and on the surface of the tongue on the other end. Therefore, it contributes at the same time to the elevation of the tongue in the velar region and to the lowering of the velum. Its role on tongue movement in speech production was not clearly demonstrated.

Four additional intrinsic muscles, completely embedded in the tongue structure, contribute to a lesser extent to the sagittal tongue shape. The *superior longitudinal* muscle shortens the tongue, and bends its blade upwards. The *inferior longitudinal* muscle depresses the tip. The *verticalis* fibers depress the tongue and flatten its surface. The *transversalis,* the only tongue muscle which fibers are transversally oriented, narrows the tongue in the coronal plane, and lengthens it in the mid-sagittal plane.

### II.2  *Muscle arrangement in the FEM structure*

Since the tongue model was essentially developed to study speech production, the proposed model only integrates the 8 muscles that are known to be largely involved in the production of basic speech movements. The palatoglossus is therefore not modeled.

#### II.2.1   The Genioglossus

The genioglossus is the largest muscle of the tongue. It is a triangular muscle running midsagittally in the central part of tongue (see figure 1). In the front this muscle inserts in the superior mental spine of the mandible. Its fibers run from the chin to the back in a fan-like manner, and the majority of them insert in the tongue body at the other extremity. In its lowest part, muscles fibers are also attached to the hyoid bone.

Functionally, this muscle influences tongue geometry in different ways, depending on the front/back location along the tongue, and each part can be independently controlled. Consequently, splitting it into three parts, a posterior one (GGp, dark grey section on Figure 1), a medium one (GGm, hartched dark grey section on Figure 1), and an anterior one (GGa, light grey section of Figure 1) is a reasonable modeling approach (Dang & Honda, 2002) [9]. The fibers of these different muscle parts join and interleave each other in the region of the insertion on the mandible (hartched light grey section on Figure 1).

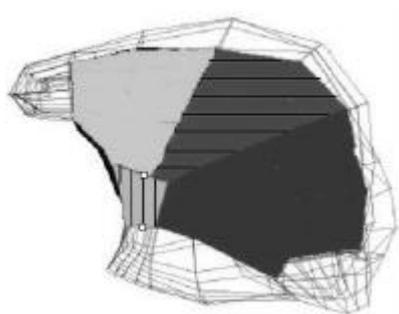

**Figure 1: Side view of Genioglossus**

The width of the GGp in the coronal plane varies from 5 mm near the superior mental spine of the mandible to 25 mm in the pharyngeal region. In the mid-sagittal plane, its maximal horizontal size is 60 mm, and its thickness is 25 mm. The GGa is located in front of and above the GGp. In the coronal plane its width varies from 5 mm near the bony insertion to 19 mm in the velar region. In the midsagittal plane, its maximal horizontal size is 37 mm and its maximal thickness is 21 mm.

In its anterior part, the limits of the genioglossus are not easy to define. Studying a foetus tongue, Langdon *et al.* (1978) [1,2] could trace fibers ending very close to the tongue tip, but Doran and Baggett (1972) [11] and Takemoto (2001) [37] did not find these fibers. According to Takemoto (2001) [37] the most anterior Genioglossus fibers end 8mm back from the tip. Based on this finding, the largest horizontal size of the GGa in the midsagittal plane is 36 mm. Its maximal vertical thickness is 32 mm, and its width in the coronal plane varies from 5 mm in the bottom to 24 mm in the palatal region.

#### II.2.2   The Hyoglossus

The Hyoglossus (HY) consists of two almost rectangular parts, located symmetrically on each side of the midsagittal plane in the posterior region of the tongue (Figure 2).

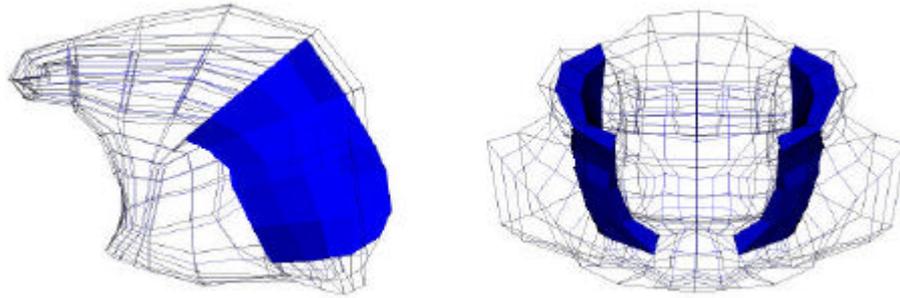

**Figure 2: Side view (left panel) and front view (right panel) of Hyoglossus**

It originates in the lateral horns of the hyoid bone and it fibers run upward along the tongue side and insert in the tongue body. It is a thin muscle of 6 mm wide in the coronal direction; in the midsagittal plane, its maximal horizontal size is 27 mm and its maximal thickness is 34 mm.

### II.2.3  The Styloglossus

The Styloglossus (ST) originates from the styloïd process, externally from the tongue. It reaches the tongue body laterally and then splits on each side into three fibers groups. The first group runs forward to the tongue tip all along the tongue sides. The second group goes inward the tongue body and joins the other inward symmetric fiber. The third one runs forward and downward toward the tongue root. All fiber groups become thinner toward their extremities.

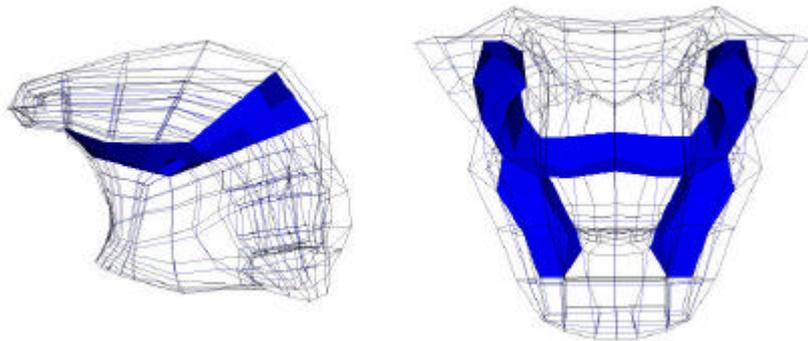

**Figure 3: Side view (left panel) and top view (right panel) of Styloglossus**

The FEM structure only accounts for the first (Figure 3, left panel) and the second (Figure 3, right panel) fiber groups.

### II.2.4  The Superior longitudinal

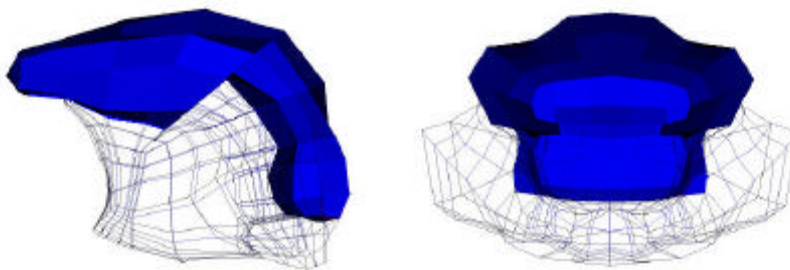

**Figure 4: Side view (left panel) and front view (right panel) of Superior longitudinal**

The superior longitudinal (SL) covers the tongue body and is located just below the mucous membrane. Clearly, its fibers insert in the apical region, but their exact extent toward the back and the sides of the tongue is still a matter of debate. According to Barnwell, Klueber and Langdon (1978) [1], some authors observed that, in the back, fibers would originate in the tongue dorsum region. On the contrary, Miyawaki [22] observed a continuity of those fibers from the low pharyngeal region to the tongue tip, and this was recently confirmed by Takemoto (2001) [37]. Takemoto also observed fibers all around the apical and palatal region, up to the insertions of the Hyoglossus. Takemoto's recent findings were taken into account in the FEM structure. Therefore, Superior Longitudinalis elements originate in the low pharyngeal region, end up in the tongue tip (Figure 4, left panel), and surround partly the whole mesh in the palatal and apical regions (Figure 4, right panel).

### II.2.5 *Inferior longitudinal*

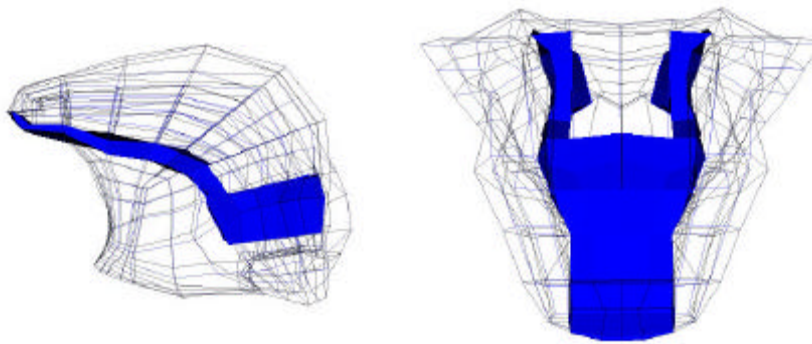

**Figure 5: Side view (left panel) and top view (right panel) of Inferior longitudinal**

Based on Takemoto's (2001) [37] observations, the elements modeling the Inferior Longitudinal (IL) in the FEM structure originate in the lower region of the tongue tip and go back to the tongue root (Figure 5, Left panel). They become thinner when they reach the tongue tip. In agreement with Barnwell et al. (1978) [2] and Miyawaki (1974) [22], these elements split into two symmetrical parts when this muscle meets the GGp (Figure 5, Right panel).

### II.2.6 *Transversalis*

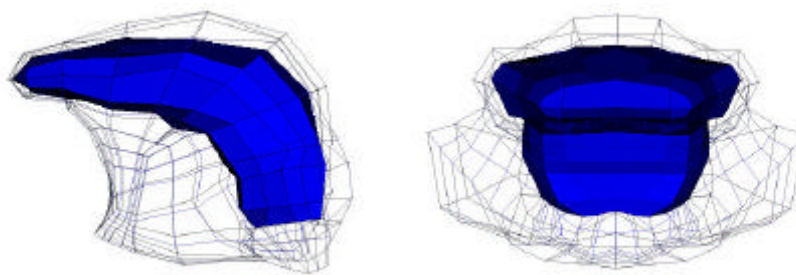

**Figure 6 Side view (left panel) and front view (right panel) of Transversalis**

The Transversalis (T) is made of thin fibers that were found in the whole tongue body, and that originate from the lingual septum in the mid-sagittal plane. They run transversally and end in the margins of the tongue. Their thickness is almost constant from the tongue tip to the back of the tongue. The FEM structure accounts for these characteristics (Figure 6).

### *II.2.7    Verticalis*

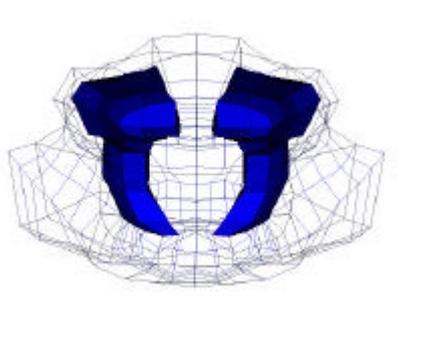

**Figure 7: Front view of Verticalis**

The Verticalis (V) is constituted of a series of thin fibers, running from the extremity of the tongue tip to the back of the tongue, and interleaving with Transversalis fibers. However, according to Miyawaki (1974) and Takemoto (2001) [37], and in contrast to the Transversalis, no Verticalis fibers were found near the mid-sagittal plane (see Figure 7)

### *II.2.8    Mylohyoid*

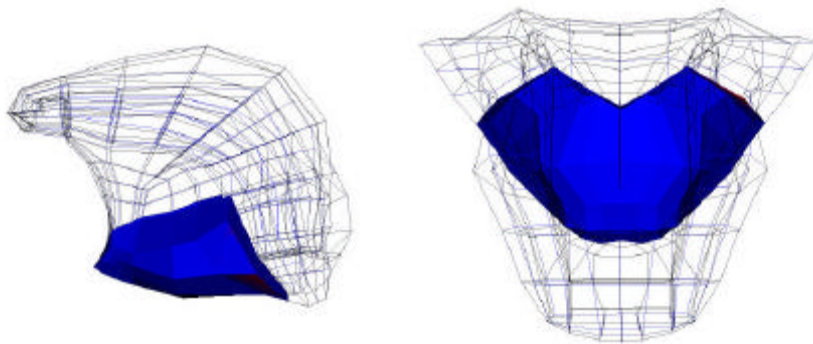

**Figure 8: Side view (left panel) and top view (right panel) of Mylohyoid**

The anatomy of the mylohyoid (M) was not discussed in Miyawaki (1974) [22] and Takemoto (2001) [37]. Hence, we used Netter's data (Netter, 1989) [24]. This muscle occupies the whole tongue floor. It is attached on one end on the inferior mental spine of the mandible and on the other end on the hyoid bone. Fibers are going medially from one side to the other (Figure 8).

### II.3  *Mesh regularization for Finite Element Computation*

Two methods are classically used to define a Finite Element mesh: manually or with automatic mesh generators.

Most of the commercial products that include automatic mesh generators use the tetrahedral meshing technique, which is by far the most common form of unstructured mesh generation (Owen, 1998) [27]. This technique is frequently based on the Delaunay criterion (Delaunay *et al.,* 1934) [8] and builds meshes that can directly be used for Finite Element Analysis.

For complex structures, as for the tongue, a manual design of the mesh is usually preferred. Hexahedral elements can be thus defined, with some advantages in comparison with tetrahedral meshes: increased accuracy, better error estimation, and quicker convergence (Craveur, 1996) [7]. Moreover, it allows a fine control of the element distribution over the mesh. In the case of tongue structure, this means for example to be able to identify elements that belong to specific muscles.

However, a manually designed mesh usually cannot be used directly for finite element analysis. Indeed, there is no *a priori* guaranty that the mesh satisfies what are called the "regularity criteria". These criteria are defined in the framework of the Finite Element method. In this framework, elasticity equations are globally solved over the domain as a summation of numerical integrations on each element of the mesh. In order to simplify the integration, a *reference element* with a regular shape (a cube in the case of hexahedral elements) is defined in a *natural* ($r, s, t$) coordinate system (Touzot *et al.* 1984 [38]; Zienkiewicz and Taylor, 1989 [41]). For each element of the deformed mesh, numerical integration is performed in the reference element, then transferred to the *actual* element using a polynomial shape function ***t*** that maps *natural* coordinates ($r, s, t$) to the *actual* coordinates of the element ($x, y, z$). The shape of an element $e$ is considered to be "regular" if it is possible, for any point inside the element, to define the function ***t*** that maps the reference element to $e$. Mathematically, this means that the Jacobian matrix $J$ of the shape function ***t*** must not be singular, i.e. that ***D****(J)*, the determinant of $J$, must never be equal to zero. A simple way to test this criterion is to calculate ***D****(J)* at each node of the element, and to check whether the computed values (eight values for a hexahedron) all keep the same sign. If Jacobian determinant value is found positive on a given node and negative on another one, this means that a point inside the element somewhere between those nodes has a zero Jacobian determinant value. The corresponding element is therefore detected as "irregular".

About 10% of the elements manually designed from the tongue Visible Human data set were detected as "irregular". The next stage of our work therefore focussed on the regularization of the tongue mesh. For this, the algorithm proposed by Luboz and colleagues (Luboz *et al.*, 2001) [20] was applied. It is based on an iterative process: nodes of irregular elements are iteratively slightly moved, until each element becomes regular. Each iteration step consists of:

1. Computing the Jacobian determinant value ***D****(J)* for each element of the mesh;
2. Detecting irregular elements (where ***D****(J)* is negative on some nodes of the element);
3. Correcting each irregular element. For this, nodes with negative ***D****(J)* values are moved in the direction of the gradient of ***D****(J)*, which is analytically computed. Displacements are weighted by the absolute values of ***D****(J)*. Displacement vectors of nodes shared by several irregular elements are the summation of displacement vectors computed for each element.

In addition, maximal nodes displacements are constrained during the iteration process, so that the corrected mesh is still close to the manually designed one. For the regularization of the Visible Human tongue mesh, the distance between the initial and final node positions cannot exceed 1mm.

The tongue mesh was successfully regularized by the algorithm, and can now be used for finite element analysis.

### III. Mechanical Modeling and differential equations

Tongue models published by Kiritani *et al.* (1976) [19], Kakita *et al.* (1985) [18], Hashimoto and Suga (1986) [16], Honda (1996) [17], or Payan and Perrier (1997) [28] all used a linear stress/strain relationship since a small deformation modelling framework was assumed. However, more recently, Napadow and colleagues (1999) [25] showed that such a small deformation framework was not perfectly adequate for tongue tissues, since tissues elongation as large as 200% (respectively 160% for contraction) were reported. Therefore, our 3D model was developed in the large deformation framework and a fair account of the tongue mechanical behavior is required.

### III.1 *Accounting for tongue muscles rheology*

Biological soft tissues are very complex materials that can feature non-linear, anisotropic and visco-elastic constitutive behaviour (Fung, 1993) [13]. Skin tissues show for example a non-linear stress-strain constitutive law, describing a quasi-incompressible and inhomogeneous material (Manschot and Brakkee, 1986 [21]; Bischoff *et al.*, 2000 [4]). Skeletal muscles show the same kind of mechanical behaviour, with specificities due to fibres orientation, i.e. with a constitutive behaviour that is not identical in the longitudinal axis and in the transverse directions (Bosboom et al., 2001 [5]; Donkelaar et al., 1999 [10]).

To our knowledge, the constitutive behaviour of human tongue has unfortunately never been experimentally measured. Since it is a complex mixture of dermis mucosa, fat tissues and muscles, its mechanical behaviour is probably very complex. Nevertheless, since most of the tongue body is composed of muscular fibres, it was decided to use for our model, in a first stage, a unique constitutive law that could be sufficiently representative of tongue muscles.. Since cardiac muscles with their interwoven fibres and their characteristics of fast muscles (REF) present a number of similarities with tongue muscles values reported by Taber and Perucchio (2000) [36] for myocardium muscles were used as input data to our mechanical modelling.

### III.2 *Mechanical modelling*

More precisely, the hypothesis of *hyperelasticity* was adopted. Whereas a material is said to be *elastic* when stress $S$ at point $X$ depends only on the values of the deformation gradient $F$, the material is said to be *hyperelastic* when stress can be derived from the deformation gradient and from a stored strain energy function $W$: $S = \frac{\partial W}{\partial E}$, where $E$ is the Lagrangian strain tensor. It is a suitable measure of deformation, since it reduces to the zero tensor when there is only rigid-body motion: $E = \frac{1}{2}(F^T F - I)$.

Due to its formulation, a hyperelastical material is said to be *path-independent*, which means that the work done by the stresses during a deformation process is dependent only on the initial and the final configurations.

The strain energy $W$ is a function of multidimensional interactions described by the nine components of $F$. It is very difficult to perform experiments to determine these interactions for any particular elastic material. Therefore, various assumptions have been made to derive simplified and realistic strain energy functions, and different formulations have been elaborated, such as the ones of the Neo-Hookean, the Ogden (1972), or the Mooney-Rivlin (1940) materials [23]. According to Fung (1993) [13], an exponential law would be particularly well adapted for biological soft tissues. Consequently, Taber and Perucchio's (2000) [36] exponential law for the myocardium seems to be adequate. Only the passive and isotropic components of the strain energy were adopted: $W_p = \frac{a}{b}(e^{bQ} - 1)$, with $Q = I_1 - 3$, where $I_1$ is the first invariant of the deformation tensor $E$: $I_1 = 3 + 2Tr(E)$.

Moreover, tongue tissues are mostly incompressible; to account for this characteristics a second term was introduced in the formula of $W$, it is *a* function of pressure $p$ acting on the tissues:

$$W = \frac{a}{b}\left(e^{bQ} - 1\right) - \frac{1}{2} p(I_3 - 1)^2$$

where $I_3$ is the third invariant of the deformation tensor $E$: $I_3 = \det(2E+I)$. This $I_3$ invariant characterizes the volume change during deformations.

In this formula, parameters *a* and *b* were chosen to replicate what is already known about tongue biomechanics (Payan & Perrier, 1997 [28]), i.e. a Young modulus close to 15kPa in the rest position (no tongue deformation), and around 110kPa when muscles are fully contracted (with a 160% deformation level, as reported by Napadow et al., 1999). The values of *a* and *b* compatible with these two constraints were respectively 2.5kPa and 0,2526. Figure 9, left panel, plots the stress/strain constitutive law that was finally modelled following the strain energy function. This law was introduced into ANSYS$^{TM}$ software in order to be approximated by a Mooney-Rivlin material. In this aim, a five parameter Mooney-Rivlin law was finally used, with $c_{10}$= 4.367kPa, $c_{01}$= -1.908kPa, $c_{20}$= 2.077kPa, $c_{11}$= -1.851kPa, $c_{02}$= 0.530kPa.

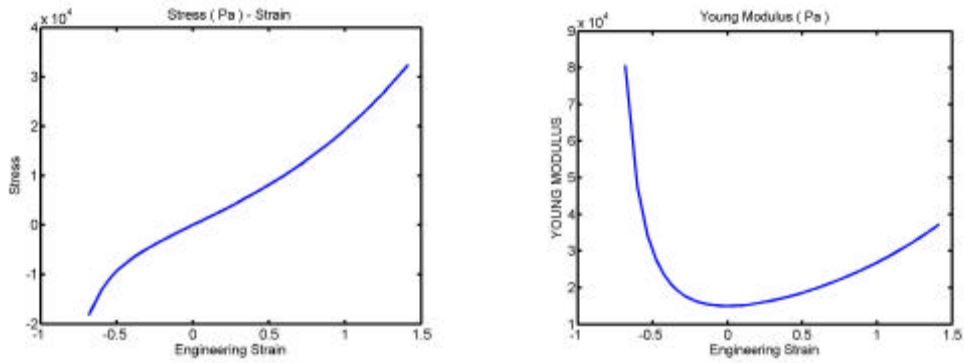

**Figure 9: Modelled stress/strain constitutive law (left panel) and its associated Young modulus/strain function ( right panel)**

The Poisson ratio value ν should be close to 0.5 in order to model nearly incompressibility of the tongue. As ν gets closer to 0.5, volume change decreases, but calculation time increases significantly. Therefore, we have chosen ν=0.49 ensuring a volume change inferior to 2% with reasonable calculation times.

### III.3 *Boundary conditions*

#### III.3.1 *Fixed points and mechanical constraints*

The tongue is highly constrained inside the mouth. Its base is attached to the mandible and to the hyoid bone, while its upper and lateral surfaces are often in contact with the palate and the teeth. At this stage of the work, the tongue model is not embedded in a 3D representation of the vocal tract. Consequently the impact onto tongue shaping of contacts and collisions with the palate and other external structures cannot be accounted for.

The current structure of the model includes a "no-displacement constraint' applied on the nodes that are attached to the mandible. In addition, nodes located at the very low limit of the tongue, i.e. in the tongue root area, are assumed to have no vertical displacements. In addition, displacements of the nodes attached to the hyoid bone are limited to the forward/backward horizontal direction. These last two constraints are a simplified way to simulate the complex and not yet modelled action of the whole set consisting of the mouth floor, the jaw opener muscles, the hyoid bone, the larynx and the external larynx muscles, which strongly limits the vertical and the lateral displacements of the tongue root. .

### *III.3.2 Functional model of the tongue muscle forces*

It was explained in Section II.2, that tongue muscle arrangement inside the FEM mesh is modelled by individual sets of adjacent elements. In addition, a main fibres direction was defined for each muscle within the associated set of element. Muscle activations were, then, functionally modelled by an external generator that applies forces on the nodes of each set of elements. The generated forces are oriented along the main fibres direction and they are mostly concentrated at the two muscle extremities in this direction. They tend to shorten the muscle when this one is activated. The main fibres orientation describes a curved path within each individual set of elements. In order to account for this factor, a distributed model of force is also applied on nodes located between the two extremities of the muscle. For each of these nodes, the distributed model exerts forces that are function of the muscle curvature, and that tends to straighten the muscle during activation (see Payan & Perrier, 1997 [28], for more details about this functional muscle model).

# IV. Simulations of movements

## IV.1 *Equations of movement*

The equation of motion, which has to be solved on each node of the FEM mesh, is defined by the following system of differential equations:

$$[M]\ddot{u}+[C]\dot{u}+[K]u=F_{int}+F_{ext},$$

where $[M]$ is the mass matrix, $[C]$ the damping matrix, $[K]$ the elasticity matrix, $F_{int}$ the internal force vector (weight and reaction forces in the model), $F_{ext}$ the external force vector and $u$ the displacement vector.

This partial differential equation is solved by ANSYS$^{TM}$ software that computes the strains, stresses and displacements of the tongue structure. $F_{ext}$ is defined by the user and corresponds here to muscles activations. A material density equal to 1 was chosen to calculate the mass matrix $[M]$. ANSYS$^{TM}$ computes automatically this matrix on the basis of each element volume and of the density. The resulting mass of the whole tongue model was thus evaluated to 160g. Weight has been implemented using the mass matrix and a gravity factor equal 9.81m.s$^{-2}$. The Mooney-Rivlin material adopted for the computation leads to the formulation of the $[K]$ matrix. The Rayleigh damping model was chosen for the definition of the $[C]$ damping matrix. This model is based on a linear combination of $[K]$ and $[M]$: $[C]=\alpha[M]+\beta[K]$, the two Rayleigh damping coefficients α and β being respectively equal to 0 and 0.028. These two values have been experimentally chosen, in order approximate a critical damping.

To solve non-linear transient equations, ANSYS$^{TM}$ uses a combination of Newton-Raphson and Newmark methods. The Newton-Raphson method typically handles non-linear static equations f({u$_n$})=0. To account for the time variation of the nodal vector u$_n$, and, then, for dynamics, the Newmark method, based on finite difference expansion in a Δt time interval, is used (Bathe, 1982):

$$\{u_{n+1}\}=\{u_n\}+\{\dot{u}_n\}\Delta t+\left[\left(\frac{1}{2}-\alpha\right)\{\ddot{u}_n\}+\alpha\{\ddot{u}_{n+1}\}\right]\Delta t^2$$

$$\{\dot{u}_{n+1}\}=\{\dot{u}_n\}+[(1-\delta)\{\ddot{u}_n\}+\delta\{\ddot{u}_{n+1}\}]\Delta t^2$$

where α and β are Newmark integration parameters, and $\dot{u}_n$ and $\ddot{u}_n$ are respectively the velocity and acceleration vectors.

Thus, at time n+1, velocity and acceleration vectors can be expressed as a function of the vectors computed at time n and of the nodal vector $\{u_{n+1}\}$ to be found. This method is unconditionally stable for $a \geq \frac{1}{4}\left(\frac{1}{2}+d\right)^2$, $d \geq \frac{1}{2}$ and $\frac{1}{2}+d+a>0$ ( Zienkiewicz ,1977 [41]).

This non-linear equation, with the unknown $\{u_{n+1}\}$ can now be solved using full Newton-Raphson method. In parallel to it, the Line search method is used, which improves new displacement vector estimation and reduces calculation time. Moreover, an adaptive descent method was activated if the problem becomes too stiff. In this case, the solver starts its calculation with a less stiff matrix (the *secant stiffness* matrix) and approaches the real solution step by step from the secant matrix to tangent matrix.

### IV.2 *Some examples of simulations*

Some examples of tongue deformation associated with specific muscle activations are given in this section. In each figure, the set of elements associated with the recruited muscle is represented in dark grey, the tongue mesh at rest (without the gravity) is represented in light grey, and the final tongue shape obtained at the end of the muscle activation is represented in white. In each simulation force muscle was a step function of 120 ms. The modelled impact of the different muscles activation is assessed on the basis of experimental measurements of vocal tract shape in the mid-sagittal plane for speech sounds, such as, for example, the X-Ray data published for French by Bothorel et al. (1986) [6] or for English by Perkell (1969) [29].

Figure 10 shows the deformation associated with a 3N force produced by the Styloglossus. The resulting tongue shape is bunched, with an elevation of the tongue dorsum in the velar region, a slightly backward displacement of the tongue body and a lowering of the tongue tip. This is typically the kind of deformation that is observed in velar sounds such as the vowel [u] and the consonant [k], which are known to be produced mainly by the recruitment of the Styloglossus. The modelled impact of this muscle seems then properly accounted in the model.

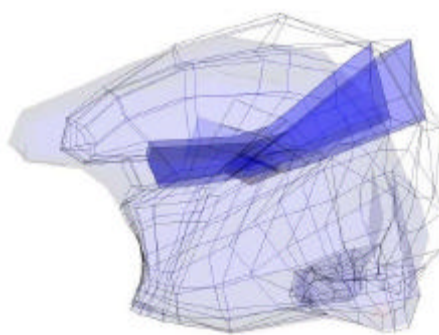

**Figure 10: Tongue deformation associated with a 3N force produced by the Styloglossus.
The temporal variation of the force is a step function of 120 ms.
Dark grey: set of elements associated with the muscle is represented in dark grey
Light grey: tongue mesh at rest (without the gravity)
White: final tongue shape at the end of muscle activation.**

Figure 11 shows the deformation associated with a 4N force produced by the Posterior Genioglossus. A strong compression of the element set corresponding to this muscle can be observed. It induces a noticeable forward movement of the back part of the tongue, associated with a slight forward displacement of the whole tongue body. In addition, a slight elevation of the upper part of the tongue is observed. The general deformation of the trend is similar to tongue shapes observed for high front vowels such as [i] that are known to be produced mainly by the activation of the Genioglossus. However, compared to the experimental observations, the elevation of the upper part of the tongue is not large enough, and the compression of the sides of the tongue in his part is too small.

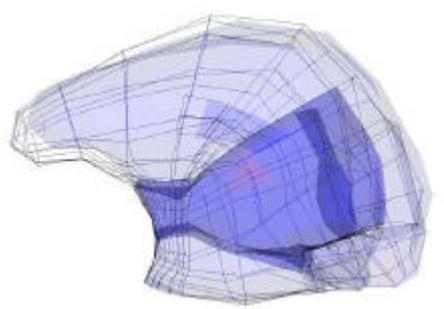

**Figure 11: Tongue deformation associated with a 3N force produced by the Posterior Genioglossus.
For details, see Figure 10**

Figure 12 shows the deformation associated with a 1N force produced by the Superior Longitudinal. A retraction and an elevation of the tongue can be observed. This is comparable to the tongue tip movement classical associated with the production of alveolar consonants such as [t]. However, experimental data for these sounds show a clear flattening of the tongue dorsum. This phenomenon is not generated by the activation of the Superior Longitudinal in the model.

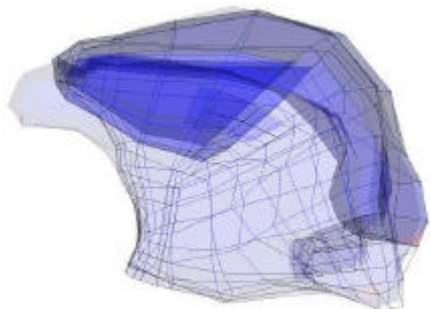

**Figure 12: Tongue deformation associated with a 1N force produced by the Superior Longitudinal.
For details, see Figure 10**

The examples presented above shows that the main features of the tongue deformation classically associated with some specific tongue muscles is fairly well accounted for by the model. However, limitations are clearly observed, which are mainly associated with an insufficient deformation of the tongue in some parts of the FEM structure that are external to the set of elements modelling the muscle. Two possible hypotheses have to be further tested in order to explain this result. The first one is related to the elastic properties of tongue tissues: the FEM structure could be too stiff in some regions, where displacements seem to be too little compared to experimental data. The second one is related to muscle control: the tongue shapes experimentally observed could be produced, by the simultaneous activation of a muscle that would induce the main deformation and of some other muscles, such as the Medium Genioglossus in the case of alveolar consonants or the Styloglossus in the case of front high vowels, that would contribute to the details of the final tongue shape. Testing thes hypotheses will involve further studies of tongue muscles and tissues rheology, associated with specific electromyographic measurements.

## V. Conclusion

A 3D biomechanical model of human tongue has been developed, in order to develop further studies aiming at testing and improving speech motor control models. In its current state, the model is not embedded in a 3D description of the vocal tract. Therefore, it cannot yet be tested in a speech production framework. However, preliminary tests of the impact of tongue muscles activation could be made. They show that the general trend is correct, since the main directions of deformation are in agreement with experimental data about speech production. However, some aspects of the modelled deformation are not in agreement with the experimental data. A more extensive evaluation will be made in the 3D space, using in particular MRI data.

Further studies will aim at refining the modelling of the elastic properties of tongue tissues, in order to clarify whether the observed discrepancies between simulations and data are due to an inadequate account of tongue biomechanics or to motor control aspects.

## Acknowledgment.

The authors would like to thank very much Professor Jacques Ohayon, from the Université de Savoie in Chambéry (France) and Professor Alain Barraud from the Institut National Polytechnique de Grenoble (France) for their valuable help to solve theoretical issues in hyperelasticity modelling and equations solving.

This work was supported by the CNRS (Poste Rouge for R. Wilhelms-Tricarico) and NIH (Grant R01 DC01925).